\documentclass[12pt]{article}

\pdfoutput = 1

\usepackage{amsmath}
\usepackage{graphicx}
\usepackage{subcaption}
\usepackage{titlesec}
\usepackage{bigints}
\usepackage{xcolor}
\usepackage{hyperref}
\usepackage{comment}
\usepackage{float}
\usepackage{relsize}

\hypersetup{colorlinks, linkcolor={blue}, citecolor={red}}
\titlelabel{\thetitle.\quad}
\numberwithin{equation}{section}

\begin{document}

\date{today}
\title{{\bf{\Large An analytical approach to compute conductivity of p-wave holographic superconductors }}}
\author{
{\bf {\normalsize Suchetana Pal}$^{a}
$\thanks{suchetanapal92@gmail.com, sp15rs004@iiserkol.ac.in}}
{\bf {\normalsize , Diganta Parai}$^{b}
$\thanks{digantaparai007@gmail.com, dp16rs028@iiserkol.ac.in}}
{\bf {\normalsize , Sunandan Gangopadhyay}$^{c}
$\thanks{sunandan.gangopadhyay@gmail.com, sunandan.gangopadhyay@bose.res.in}}\\
$^{a,b}${\normalsize\textit{Department of Physical Sciences,}}\\
{\normalsize\textit{Indian Institute of Science Education and Research Kolkata,}}\\
{\normalsize\textit{Mohanpur, Nadia, West Bengal, 741 246, India.}}\\
$^{c}${\normalsize\textit{Department of Theoretical Sciences,}}\\
{\normalsize\textit{S. N. Bose National Centre for Basic Sciences,}}\\
{\normalsize\textit{Block - JD, Sector - III, Salt Lake, Kolkata - 700 106, India.}}}
\date{}

\maketitle

\begin{abstract}
\noindent In this article we have analytically deduced the frequency dependent expression of conductivity and the band gap energy in $AdS_{4}$ Schwarzschild background for p-wave holographic superconductors considering Einstein-Yang-Mills theory. We also used the self consistent approach to obtain the expressions of conductivity for different frequency ranges at low temperature. We then compared the imaginary part of conductivity at low frequency region. The band gap energy obtained from these two methods seem to agree very well. 
\end{abstract}

\section{Introduction}

The AdS/CFT correspondence proposed by Maldacena \cite{maldacena1999large,aharony2000large,witten1998anti1,witten1998anti2,gubser1998gauge} states that a weakly coupled gravity theory in $AdS_{n+1}$ spacetime is equivalent to a strongly coupled conformal field theory $CFT_{n}$ in one less dimension \cite{natsuume2015ads,hartnoll2018holographic}. This conjecture derived from string theory has been used extensively to understand the strongly coupled phenomena in field theories by looking at a weakly coupled dual gravity theory. A very interesting application of this correspondence is the construction of holographic superconductors. The term ``holographic" implies, by looking at a two (spatial) dimensional superconductor one can identify a three-dimensional image that consists of a charged black hole with non-trivial hair \cite{papantonopoulos2011gravity}. These holographic superconductors can successfully reproduce many important properties of high $T_c$ superconductors. The second-order superconducting phase transition below a certain critical temperature can be understood by the condensation of a charged scalar field that leads to $U(1)$ symmetry breaking near the black hole horizon in the dual gravitational description. There exist a large number of studies to describe the Meissner effect which is another important feature of superconductors from the holographic superconductor point of view \cite{albash2008holographic,gangopadhyay2014holographic, pal2018noncommutative,pal2020meissner}. There are also studies on effect of nonlinear electrodynamics and higher curvature correction on holographic superconductor \cite{sunandan2012analytic1,gangopadhyay2012analytic1}. Different mechanisms have been proposed where at a finite temperature the system undergoes a spontaneous symmetry breaking and enters the superconducting phase
\cite{gubser2008gravity,gubser2005phase,gubser2008breaking,hartnoll2008building, gubser2008colorful}. In this paper, we consider a p-wave holographic superconductor model as considered in \cite{gubser2008gravity,gubser2008colorful,akhavan2011p,gangopadhyay2012analytic} and try to shed light on the optical conductivity at the superconducting phase in an analytical approach.

 Our analysis is based on the simplest example of p-wave holographic superconductor depicted by $SU(2)$ Einstein-Yang-Mills theory given by the following action
\begin{equation}
\label{action}
 \mathcal{S}= \int d^{d}x~\sqrt{-g}~\Big[\frac{1}{2}(R-2\Lambda)-\frac{1}{4}F^{a}_{\mu\nu}F^{a\mu\nu}\Big],   
\end{equation}
where the negative cosmological constant  $\Lambda=-\frac{(d-1)(d-2)}{2L^2}$ with $L=1$ and Yang-Mills field strength $F_{\mu\nu}^{a}=\partial_{\mu}A^{a}_{\nu}-\partial_{\nu}A_{\mu}^{a}+qf^{abc}A_{\mu}^{b}A_{\nu}^{c}$. The gauge field can be written as $A=A^a_{\mu}\sigma^{a}dx^{\mu}$, where $\sigma^a$ are the generators of the $SU(2)$ group and $(a,b,c)= (1,2,3)$ are the indices of the generators.
The equation of motion for the field variable $A_{\mu}$ coming from the above action is given by 
\begin{equation}
\label{eom1}
\frac{1}{\sqrt{-g}}\partial_{\mu}(\sqrt{-g}F^{a\mu\nu}) + qf^{abc}A^{b}_{\mu} F^{c\mu\nu}=0.   
\end{equation}

\noindent  For finite $q$, one must consider the effect of the gauge field on the metric but when $q$ is large the back reaction is negligible. This $q\rightarrow\infty$ limit is known as probe limit which simplifies the problem but retains important properties of the system. In this paper our analysis is done in the probe limit.   

 In this paper we aim to obtain the frequency dependent expression for conductivity in $AdS_{4}$ Schwarzschild background.  The paper is organized in seven sections. We start with the basic formalism in section 2 where field equations are discussed. In section 3 we provide the relationship between critical temperature $T_c$ and charge density $\rho$. In section 4 we analyse the system at low temperature limit ($T\rightarrow0$) and  determine the behaviour of the charged field $\psi$ and gauge field $\phi$, that is consistent with the boundary conditions and also obtain the relationship between condensation operator $\langle \mathcal{O}_{1}\rangle$ and critical temperature $T_{c}$. In section 5 we we discuss conductivity and compute band gap energy for the case $\psi_0=0$. In section 6 we consider the case $\psi_1=0$ and compute conductivity \textbf{and band gap energy}. In section 7 we draw conclusions from the findings.


\section {Discussion of the holographic model}
For this part of our analysis we have chosen the fixed background of $3+1$-dimensional Schwarzschild $AdS$ black hole, whose metric reads
\begin{equation}
\label{metric}
ds^2=-f(r)dt^2+\frac{1}{f(r)}dr^2+r^2(dx^2 + dy^2)
\end{equation}
where,
\begin{eqnarray}
\label{fr}
f(r)=r^2g(r)~, ~g(r)=\left(1-\frac{r_+^3}{r^3}\right)~.
\end{eqnarray}
Here $r_+$ is the horizon radius. The Hawking temperature is given by 
\begin{equation}
T=\frac{3r_+}{4\pi}~.
\label{htemp}
\end{equation}

\noindent In order to investigate the metal/superconductor phase transition let us consider the following ansatz
\begin{equation}
\label{ansatz1}
A=\phi(r)\sigma^{3}dt+\psi(r)\sigma^{1}dx~.
\end{equation}
\noindent Here the gauge field $A^3_t=\phi(r)$ is the $U(1)$ subgroup of $SU(2)$ and is associated with the chemical potential in the boundary field theory. The charged field $A^1_x=\psi(r)$ is associated with the condensation operator $\langle\mathcal{O}\rangle$ in the boundary field theory whose condensation is responsible for $U(1)$ symmetry breaking. Note that our analysis has been done in the probe limit, hence we did not consider the back reaction of the gauge field on the metric eq.[\ref{fr}].

\noindent Plugging on the ansatz given by eq.[\ref{ansatz1}] in eq.[\ref{eom1}] we obtain the following equations of motion for the field variables $\phi(z)$ and $\psi(z)$ respectively 
\begin{equation}
\label{eomphi}
\phi''(z)-\frac{\psi^2(z)}{r_+^2g(z)}\phi(z)=0~,
\end{equation}

\begin{equation}
\label{eompsi}
\psi''(z)+\frac{g'(z)}{g(z)}\psi'(z)+\frac{\phi^2(z)\psi(z)}{r_+^2g^2(z)}=0~.
\end{equation}
Note that here we  considered the coordinate change $z=\frac{r}{r_+}$ for simplicity. At the horizon($r=r_+$) $z=1$ and at the boundary($r\rightarrow\infty$) $z\rightarrow0$. 

\noindent Let us now discuss the boundary conditions. At the horizon  $\phi(1)=0$ and $\psi(1)$ is finite. At the boundary, that is $z\rightarrow0$ the behaviour of $\phi(z)$ and $\psi(z)$ are as follows
\begin{equation}
\label{bc1}
\phi(z)=\mu-\frac{\rho}{r_+}z~,
\end{equation}

\begin{equation}
\label{bc2}
\psi(z)=\psi_0+\frac{\psi_1}{r_+}z~.
\end{equation}
According to the AdS/CFT dictionary, $\mu$ and $\rho$ respectively represent the dual to chemical potential and charge density at the boundary. $\psi_{0}$ and $\psi_{1}$ are related to the source and the expectation value of the condensation operator.

\noindent For now let us consider $\psi_0=0$. We will now discuss this case in detail. Keeping in mind the behaviour at the boundary,  we may write 
\begin{equation}
\label{psi1}
\psi(z)=\frac{\langle\mathcal{O}_{1}\rangle}{\sqrt{2}r_+}z F(z)~.
\end{equation}

\noindent In section 6, we also briefly discuss the case where we set $\psi_1=0$. In this case, we write  

\begin{equation}
\label{psi0}
\psi(z)=\frac{\langle\mathcal{O}\rangle}{\sqrt{2}}F(z)~.
\end{equation} 

\noindent For both of these cases $F(z)$ obeys the following conditions

\begin{eqnarray}
F(0)= 1, ~~~~~~~~~ F'(0)=0~.
\label{F(0)}
\end{eqnarray}

\section{Relation between critical temperature $T_c$ and charge density$\rho$}
In this section, our analysis will be mostly focused around $T\rightarrow T_c$ and we will develop the relationship between $T_c$ and $\rho$, which is necessary for later analysis and future constructions. But we will not be discussing this in great detail as it is already available in the literature \cite{gangopadhyay2012analytic}. 

\noindent Using the fact that at the critical temperature $T_c$, $\psi(z)=0$ from eq.[\ref{eomphi}] we obtain

\begin{equation}
\phi(z)= \lambda r_{+(c)}(1-z),~~~~~~\lambda= \frac{\rho}{r_{+(c)}^2}~.
\label{phiTc}
\end{equation}

\noindent By substituting $\phi(z)$ and $\psi(z)$ from eq.[\ref{phiTc}] and eq.[\ref{psi1}] respectively in eq.[\ref{eompsi}], we get

\begin{equation}
F''(z)-\left(\frac{3z^2}{1-z^3}-\frac{2}{z}\right)F'(z)-\left(\frac{3z}{1-z^3}\right)F(z)+\frac{\lambda^2}{(1+z+z^2)^2}F(z)=0~.
\label{eqFzTc}
\end{equation}
 \noindent This equation can be recast in Sturm-Liouville form corresponding to eigenvalue $\lambda^2$, which minimizes the following expression
 
 \begin{equation}
\lambda^2= \frac{\int_{0}^{1} dz [z^2(1-z^3)F'(z)^2+3z^3F(z)^2]}{\int_{0}^{1} dz \frac{z^2(1-z)}{1+z+z^2}F(z)^2} ~.    
 \end{equation}
 Here we choose the trial function $F_\beta(z)=1-\beta z^2$, which obeys the conditions given by eq.[\ref{F(0)}]. The minimum $\lambda^2$ is attained for $\beta=0.5078$. 
 
 \noindent These findings yield \cite{gangopadhyay2012analytic} 
 \begin{equation}
T_c= \frac{3}{4\pi}r_{+(c)}=\frac{3}{4\pi}\sqrt{\frac{\rho}{\lambda_{\beta}}} \approx 0.1239\sqrt{\rho}~. 
\label{rhoTc}
 \end{equation}

\section {Condensation operator $\langle\mathcal{O}_{1}\rangle$ at low temperature}
As we are interested in low temperature limit ($T\rightarrow0$), we consider the scaling $z=\frac{s}{b}$ where $b\rightarrow\infty$. We will determine $b$ later on, and it will be clear that $b\rightarrow \infty $ corresponds to low temperature limit \cite{siopsis2010analytic,ghorai2018conductivity}. Under this condition the dominant contribution comes from neighbouring region of the boundary ($z\rightarrow0$) and eq.[\ref{eomphi}] and eq.[\ref{eompsi}] takes the following forms respectively  

\begin{equation}
\label{eqphis1}
\phi''(s)-\frac{\langle\mathcal{O}_{1}\rangle^2}{2 r_+^4 b^4 } s^2 F^2(s)\phi(s)=0~,
\end{equation}

\begin{equation}
\label{eqFs1}
F''(s)+\frac{2}{s}F'(s)+\frac{\phi^2(s)}{r_+^2 b^2} F(s)=0~.
\end{equation}

\noindent  We aim to obtain the solutions of the eq.[\ref{eqphis1}] and eq.[\ref{eqFs1}] iteratively, which are consistent with the boundary conditions. To do that, we will start with the following form of $F(s)$, which is essentially the behavior of $F(s)$ at $s> 1$ or $z>\frac{1}{b}$

\begin{equation}
\label{Fs1}
F(s)\approx\frac{\alpha}{s}~.
\end{equation}

\noindent Here $\alpha$ is a constant to be determined later.
Substituting $F(s)$ from eq.[\ref{Fs1}] in from eq.[\ref{eqphis1}], we get 
\begin{equation}
\label{phis}
\phi(s)= C_1 e^{-s}+C_2 e^{s}.
\end{equation}

\noindent Now we choose $b$ as 
\begin{equation}
b=\frac{\sqrt{\langle\mathcal{O}_{1}\rangle \alpha}}{2^{\frac{1}{4}}r_+}.
\label{b}   
\end{equation}
\noindent Using the condition $\phi(z)=0$ at the horizon ($z\rightarrow 1$) for all values of $b$, it is easy to show that $C_2 = 0$ and hence $\phi(s)$ can be written as 
\begin{equation}
\label{phis1}
\phi(s)= C r_+ b e^{-s}
\end{equation}

\noindent where, $C_1= \frac{C\sqrt{\langle\mathcal{O}_{1}\rangle \alpha}}{2^{\frac{1}{4}}}$. 

\noindent Eq.[\ref{b}] shows that $b\rightarrow \infty$ as $r_+\rightarrow 0$, and hence from eq.[\ref{htemp}] it is easy to note that it corresponds to low temperature limit as we claimed earlier.

\noindent Now we proceed to estimate a more accurate behaviour of $F(s)$ that is consistent with the conditions $F(z)=1$ and $F'(z)=0$ at the boundary $z=0$. By  substituting $\phi(s)$ from eq.[\ref{phis1}] in [\ref{eqFs1}], we obtain
\begin{equation}
\label{eqFs2}
F''(s)+\frac{2}{s}F'(s)+C^2  e^{-2s} F(s)=0~.
\end{equation}

\noindent The above equation can be solved using the Sturm-Liouville approach in the interval $(0,\infty)$. The corresponding eigenvalue $C^2$ is given by

\begin{equation}
\label{evalue}
C^2=\frac{\int_{0}^{\infty}s^2 F'(s)^2 ds}{\int_{0}^{\infty}s^2 e^{-2s}F^2(s) ds}~.
\end{equation}

\noindent Here we choose the following trial function as the eigen function $F(s)$ that minimizes
eq.[\ref{evalue}] and is consistent with the boundary conditions
\begin{equation}
\label{trialfnF1}
F(s)= \frac{\alpha}{s} \tanh{\frac{s}{\alpha}}.
\end{equation}

\noindent We obtain the minimum for $\alpha = \alpha_{S.L}= 0.8179$ and that corresponds to $C= C_{S.L}=2.4065$.

\noindent Next we aim to solve eq.[\ref{eqphis1}] once again perturbatively by substituting $F(s)$ from eq.[\ref{trialfnF1}] and considering $\phi(s)$ from eq.[\ref{phis1}] as the zeroth order solution. We obtain

\begin{equation}
\label{phis2}
\phi(z)=c_{1}-c_{2}\frac{bz}{\alpha}+ b r_+ C e^{-bz}\Bigg[1-\bigg(\frac{2\alpha}{\alpha+2}\bigg)^2 e^{-\frac{2bz}{\alpha}} {}_3F_{2}\left\{2,1+\frac{\alpha}{2},1+\frac{\alpha}{2}; 2+\frac{\alpha}{2},2+\frac{\alpha}{2};  -e^{-\frac{2bz}{\alpha}}\right\}\Bigg]~.
\end{equation}

\noindent Using the fact that at $z=1$, $\phi(z)=0$ and $b\rightarrow \infty$, we may show that $c_{1}=c_{2}=0$. Finally, we obtain

\begin{equation}
\label{phis3}
\phi(z)=b r_+ C e^{-bz}\Bigg[1-\bigg(\frac{2\alpha}{\alpha+2}\bigg)^2 e^{-\frac{2bz}{\alpha}} {}_3F_{2}\left\{2,1+\frac{\alpha}{2},1+\frac{\alpha}{2}; 2+\frac{\alpha}{2},2+\frac{\alpha}{2};  -e^{-\frac{2bz}{\alpha}}\right\}\Bigg]~.
\end{equation}

\noindent By comparing the coefficient of $z$ from the above equation with the boundary behaviour of $\phi(z)$ given by eq.[\ref{bc1}], we obtain
\begin{equation}
\label{roh/r+}
\frac{\rho}{r_+}= 0.4911 b^2 r_+ C~.
\end{equation}

\noindent Now we substitute $\rho$ from eq.[\ref{rhoTc}] and $b$ from eq.[\ref{b}] in eq.(\ref{roh/r+}) and get

\begin{equation}
\label{O-Tc1}
\sqrt{\langle\mathcal{O}_{1}\rangle}=9.7624 T_c\equiv\xi T_{c}~.
\end{equation}

\noindent Restoring the $z$ coordinate now, we may write $F(z)$ ans $\psi(z)$ as following
\begin{eqnarray}
\label{FzPsiz1}
F(z)= \frac{\alpha}{bz} \tanh{\frac{bz}{\alpha}},\\
\label{psifinal}
\psi(z)=\frac{\langle\mathcal{O}_{1}\rangle}{\sqrt{2}r_+}\frac{\alpha}{b} \tanh{\frac{bz}{\alpha}}.
\end{eqnarray}

\noindent Interestingly we can determine the constant $C$ and $\alpha$ by direct analytical approach as eq.[\ref{eqFs2}] is analytically solvable. Using the condition $F(0)=1$ from eq.[\ref{eqFs2}], we get

\begin{equation}
\label{F2}
F(s)= \frac{\pi}{2s}\bigg[Y_0(C)J_{0}(Ce^{-s})-J_{0}(C)Y_{0}(Ce^{-s})\bigg]~.
\end{equation}

\noindent To compute the constant $C$ we will use the condition $F(s)\rightarrow0$ as $s\rightarrow\infty$. From eq.[\ref{F2}], we thus obtain

\begin{equation}
\label{besselJ}
J_{0}(C)= 0~.
\end{equation}

\noindent The above equation implies $C= C_{direct}=2.4048$,  which is the first root of the Bessel function $J_{0}$. We can see that this result agrees well with our previous estimate of the constant $C$ obtained using the Sturm-Liouville approach. Now $F(s)$ may be written as follows

\begin{equation}
\label{Fs3}
F(s)= \frac{\pi}{2s}\bigg[Y_0(C)J_{0}(Ce^{-s})\bigg]~.
\end{equation}

\noindent Now let us compare the behaviour of the $F(s)$ from the above eq.[\ref{Fs3}] with eq.[\ref{Fs1}] for $s> 1$ or, $z>\frac{1}{b}$ to find out $\alpha$. This gives

\begin{eqnarray}
\label{Fs4}
F(s)= \frac{\pi Y_0(C)}{2s}\\
\alpha= \frac{\pi Y_0(C)}{2} = 0.8009~.
\end{eqnarray}

\noindent We see that $\alpha=\alpha_{direct}=0.8009$ is in good agreement with the previous estimate of $\alpha_{S.L}$~.

\noindent Following the earlier steps using $F(s)$ given by eq.[\ref{Fs3}], and considering $\phi(s)$ from eq.[\ref{phis1}] as the zeroth order solution we compute $\phi(z)$ from eq.[\ref{eqphis1}] and the relationship between  $\langle   \mathcal{O}_{1}\rangle$ and $T_c$.

\begin{equation}
\label{phis4}
\phi(z)= b r_+ C e^{-bz} {}_3F_{4}\left\{\frac{1}{2},\frac{1}{2},\frac{1}{2} ; 1,1,\frac{3}{2},\frac{3}{2};  -C^2 e^{-2bz}\right\}
\end{equation}

\begin{equation}
\label{O-Tc}
\sqrt{\langle\mathcal{O}_{1}\rangle}=10.0518 T_c~.
\end{equation}

\noindent In this case we may write $F(z)$ and $\psi(z)$ as following 
\begin{eqnarray}
\label{FzPsiz2}
F(z)= \frac{\pi}{2bz}\bigg[Y_0(C)J_{0}(Ce^{-bz})\bigg],\\
\psi(z)=\frac{\langle\mathcal{O}_{1}\rangle}{\sqrt{2}r_+} \frac{\pi}{2b}\bigg[Y_0(C)J_{0}(Ce^{-bz})\bigg].
\end{eqnarray}

\noindent It is worth noting that for both the cases given by eq.[\ref{FzPsiz1}] and eq.[\ref{FzPsiz2}], it is possible to write $F(z)= 1+ \mathcal{O}(z^2)$ as expected.

\begin{figure}[h!]
	\centering
	\includegraphics[width=11cm]{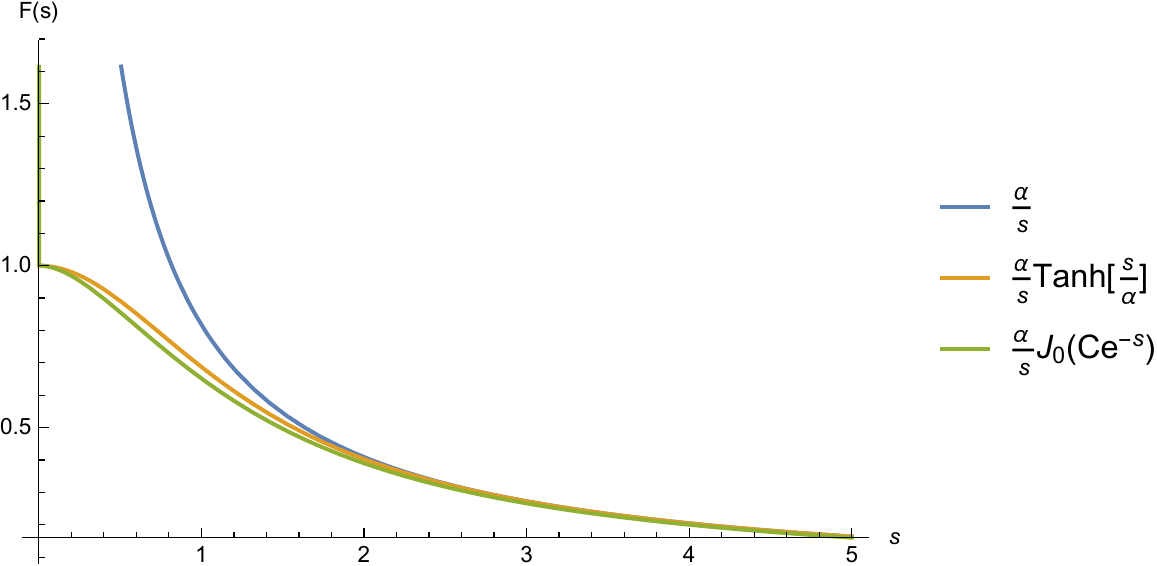}
	
	\caption{$F(s)$ vs. $s$}
	\label{psifig}
\end{figure}

\noindent To get a clear idea about the behaviour of $F(s)$ obtained from two different methods, in Figure[\ref{psifig}] we have plotted $F(s)$ vs. $s$. The blue, orange and green curves represent eq.[\ref{Fs1}], [\ref{trialfnF1}] and [\ref{Fs3}] respectively. By looking at blue curve representing eq.[\ref{Fs1}] we can easily say that indeed it is the behaviour of $F(s)$ at $s> 1$ or, $z>\frac{1}{b}$ and by looking at the orange and green curve it is clear that the behaviour of $F(s)$ predicted by eq.[\ref{trialfnF1}] and [\ref{Fs3}] are very similar as expected.

\begin{figure}[h!]
	\centering
	\includegraphics[width=16cm]{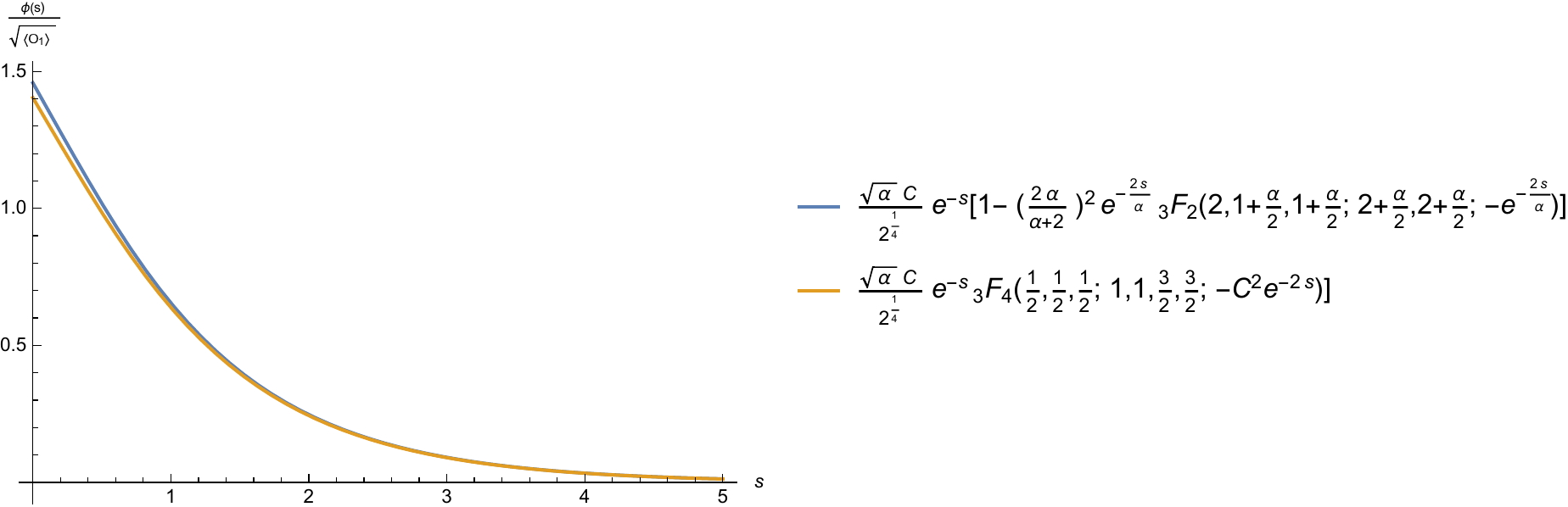}
	
	\caption{$\frac{\phi(s)}{\sqrt{\langle\mathcal{O}_{1}\rangle}}$ vs. $s$}
	\label{phifig}
\end{figure}

\noindent In Figure[\ref{phifig}], we have plotted $\frac{\phi(s)}{\sqrt{\langle\mathcal{O}_{1}\rangle}}$ vs. $s$. The blue and orange curves respectively represents eq.[\ref{phis3}] and [\ref{phis4}] which suggests the behaviour of $\phi$ depicted by these equations are very similar. Note that for the rest of the analysis we will use $\alpha=\alpha_{S.L}=0.8179$~.

\begin{table}[ht]
   \caption{$\alpha$, $C$ and $\sqrt{\langle\mathcal{O}_{1}\rangle}$  obtained from Sturm-liouville approach and direct analytical approach: }   
   \centering                        \begin{tabular}{|c| c| c| c| } \hline
   Approach          & $\alpha$ &   $C$    & $\sqrt{\langle\mathcal{O}_{1}\rangle}$  \\
   \hline
   S.L               & 0.8179   &  2.4065  &  10.0518 $T_c$      \\ 
   \hline
    Direct          &  0.8009   &  2.4048  &  9.7624  $T_c$      \\ 
   \hline
   \end{tabular}
   \label{t1}  
 \end{table}

\section{Conductivity at low temperature}
\noindent To study the conductivity at the boundary, we will consider an electromagnetic perturbation in the bulk by applying a non-zero gauge field in the $y$- direction. Let us consider the following ansatz  

\begin{equation}
\label{ansatz2}
A_y=A(r)e^{-i\omega t}\sigma^{3} ~.
\end{equation}

\noindent Note that, as our analysis is done on probe limit we do not consider the back reaction of this newly introduced component ($A_y$) on the metric or the other components of the gauge field ($A_t$, $A_x$). Plugging in the above ansatz in the equation of motion given by eq.[\ref{eom1}], we get
 
\begin{equation}
    A^{\prime\prime}(r)+\frac{f^{\prime}(r)}{f(r)}A^{\prime}(r)+\left[\frac{\omega^2}{f^2(r)}-\frac{\psi^2(r)}{r^2f(r)}\right]A(r)=0~.
    \label{eomAr}
\end{equation}

\noindent Switching to $z$ coordinate, we obtain   
\begin{equation}
\label{eomA}
A''(z)+\frac{g'(z)}{g(z)}A'(z)+ \frac{1}{r_+^2}\Big[\frac{\omega^2}{g^2(z)}-\frac{\psi^2(z)}{g(z)}\Big]A(z)=0.
\end{equation}

\noindent At the boundary the behaviour of the gauge field $A(z)$  can be found from eq.(\ref{eomAr}) given by 
\begin{equation}
\label{bc3}
A(z)=A_0+\frac{A_1}{r_+}z.
\end{equation}

\noindent The conductivity $\sigma_{yy}$ for our system is given as following (refer Appendix) \cite{srivastav2020p,mohammadi2019conductivity,sheykhi2018holographic}
\begin{equation}
\label{sigma}
\sigma_{yy}=-\frac{iA_1}{\omega A_0}=-\frac{ir_+A'(0)}{\omega A(0)}~.
\end{equation}
 
\noindent Let us now switch to the tortoise coordinate (as $z\rightarrow1$, $r_*\rightarrow-\infty$) defined as follows, where the integration constant is chosen such that at the boundary ($z=0$), $r_\ast=0$
\begin{equation}
\label{tor}
r_{\ast}=\int \frac{dr}{f(r)}=\frac{1}{6r_+}[2\ln(1-z)-\ln(1+z+z^2)-2\sqrt{3}\tan^{-1}\frac{\sqrt{3}z}{2+z}].
\end{equation}
\noindent Now eq.[\ref{eomA}] takes the following form
\begin{equation}
\label{shm1}
A''(r_{\ast})+\omega^2 A(r_{\ast})= VA(r_\ast) ,~~~~~ V(r)= \frac{\psi^2(r) f(r)}{r^2}.
\end{equation}

\noindent Note that at the horizon ($r=r_+$), $V=0$. Taking into account the ingoing boundary condition \cite{siopsis2010analytic} at the horizon, solving the above equation, we get
\begin{equation}
\label{A1}
A~\sim e^{-i\omega r_\ast}\sim (1-z)^{-\frac{i\omega}{3r_+}}.
\end{equation}
\noindent Near the horizon ($z=1$), the main contribution in $r_*$ comes from the first term as given in eq.[\ref{tor}].

\noindent In order to obtain an expression for $A(z)$, eq.[\ref{eomA}] ought to be solved taking into account the boundary behaviour. We may now write 
\begin{equation}
\label{A2}
A(z)=(1-z)^{-\frac{i\omega}{3r_+}} G(z)~.
\end{equation}

\noindent Substituting $A(z)$ from eq.[\ref{A2}] in eq.[\ref{eomA}], we get

\begin{eqnarray}
\label{Gz1}
3(1-z^3)G''(z)&-&\left[9z^2 - 2(1+z+z^2)\frac{i\omega}{r_+}\right]G'(z)\nonumber\\&-&\left[\frac{3\psi^2(z)}{r_+^2}-(1+2z)\frac{i\omega}{r_+}-\frac{(2+z)(4+z+z^2)}{3(1+z+z^2)}\frac{\omega^2}{r_+^2}\right] G(z)=0 ~.
\end{eqnarray}

\noindent At the horizon $(z=1)$, from above equation we deduce 
\begin{equation}
\bigg[3-\frac{2i\omega}{r_+}\bigg]G'(1)+\bigg[\frac{\psi^2(1)}{r_+^2}-\frac{i\omega}{r_+}-\frac{2\omega^2}{3r_+^2}\bigg]G(1)=0 ~.
\label{G(1)-G'(1)}
\end{equation}

\noindent Turning on the low temperature limit, eq.[\ref{Gz1}] may be approximated as 

\begin{eqnarray}
\label{Gz2}
G''(z)
+ \frac{2i\omega}{3r_+} G'(z) -\bigg[\frac{\psi^2(z)}{r_+^2}-\frac{i\omega}{3r_+}+\frac{8\omega^2}{9r_+^2}\bigg]G(z)=0.
\end{eqnarray}

\noindent Substituting $\psi(z)$ from eq.[\ref{psifinal}], we exactly solve the above equation and obtain 

\begin{equation}
\label{Gz}
G(z)= e^{-\frac{i\omega}{3r_+}z}\bigg[c_+ P^{\alpha\sqrt{1-\frac{\sqrt{2}\omega^2}{\alpha\langle\mathcal{O}_{1}\rangle}}}_{\frac{1}{2}(-1+\sqrt{1+4\alpha^2})} \bigg(\tanh\frac{bz}{\alpha}\bigg)+c_- P^{-\alpha\sqrt{1-\frac{\sqrt{2}\omega^2}{\alpha\langle\mathcal{O}_{1}\rangle}}}_{\frac{1}{2}(-1+\sqrt{1+4\alpha^2})} \bigg(\tanh\frac{bz}{\alpha}\bigg)\bigg],
\end{equation}

\noindent  where $P^{\mu}_{\nu} $ are the fractional Legendre functions. Finally we may write $A(z)$ for low frequency ($\omega<< \langle\mathcal{O}_{1}\rangle$) region as

\begin{equation}
A(z)=(1-z)^{-\frac{i\omega}{3r_+}} e^{-\frac{i\omega}{3r_+}z}\bigg[c_+ P^{\alpha}_{\frac{1}{2}(-1+\sqrt{1+4\alpha^2})} \bigg(\tanh\frac{bz}{\alpha}\bigg)+c_- P^{-\alpha}_{\frac{1}{2}(-1+\sqrt{1+4\alpha^2})} \bigg(\tanh\frac{bz}{\alpha}\bigg)\bigg],
\end{equation}

\noindent Using the definition of conductivity from eq.[\ref{sigma}] for low temperature and low frequency, we write

\begin{equation}
\label{sigmaC}
\sigma(\omega)= 0.4616 i \frac{\sqrt{\langle\mathcal{O}_{1}\rangle}}{\omega} \bigg(\frac{1-1.3911\frac{c_+}{c_-}}{1-0.4085\frac{c_+}{c_-}}\Bigg).
\end{equation}

\noindent Next, we aim to determine the ratio $\frac{c_+}{c_-}$ . Note that at $z\rightarrow1$, $\tanh(\frac{bz}{\alpha})\approx1$, and under this condition we may approximate 

\begin{equation}
P^{\pm\alpha}_{\frac{1}{2}(-1+\sqrt{1+4\alpha^2})} \bigg(\tanh\frac{bz}{\alpha}\bigg)= \frac{2^{\pm\frac{\alpha}{2}}}{\Gamma(1\mp \alpha)}\bigg(1-\tanh\frac{bz}{\alpha}\bigg)^{\mp
\frac{\alpha}{2}}+...~~.
\label{identity}
\end{equation}

\noindent Now for low frequency region from eq.[\ref{Gz}], we also get

\begin{eqnarray}
G(1)=\bigg[\frac{c_+}{\Gamma(1-\alpha)}e^b+\frac{c_-}{\Gamma(1+\alpha)}e^{-b}\bigg]e^{-\frac{i\omega}{3r_+}},~G'(1)=\bigg[\frac{c_+(b-\frac{i\omega}{3r_+})}{\Gamma(1-\alpha)}e^b+\frac{c_-(b+\frac{i\omega}{3r_+})}{\Gamma(1+\alpha)}e^{-b}\bigg]e^{-\frac{i\omega}{3r_+}}~.
\label{G(1)}
\end{eqnarray}
\noindent Using eq.[\ref{G(1)-G'(1)}] and eq.[\ref{G(1)}]  the ratio $\frac{c_+}{c_-}$ becomes
\begin{equation}
\label{ratio}
\frac{c_+}{c_-}= -e^{-2b}\frac{\Gamma(1-\alpha)}{\Gamma(1+\alpha)}\bigg[\frac{b-3}{b+3}+\frac{4(b^2-3)}{b(b+3)^2}\frac{i\omega}{r_+}+\mathcal{O}(\omega^2)\bigg]~.
\end{equation}

\noindent Substituting the above ratio in eq.[\ref{sigmaC}], we obtain $\sigma(\omega)$ at low frequency ($\omega\rightarrow 0$). This yields the following equations 

\begin{eqnarray}
Im [\sigma(\omega)]\approx0.4616\frac{\sqrt{\langle\mathcal{O}_{1}\rangle}}{\omega},
\label{imomega1}\\ 
Re[\sigma(\omega=0)]\sim e^{-2b}\big[1+\mathcal{O}(1/b)\big]\equiv e^\frac{{-E_{g}}}{T},
\label{gap} \\
E_g=\frac{3\sqrt{\alpha  \langle\mathcal{O}_{1}\rangle}}{2^{\frac{5}{4}}\pi}\approx0.3631\sqrt{\langle\mathcal{O}_{1}\rangle}~.
\label{bandgap}
\end{eqnarray}

\noindent We have used $\alpha= \alpha_{SL}=0.8179$ obtained from the SL method in section 4 in the above equation.

Eq.[\ref{gap}] depicts that the zero frequency limit of $Re[\sigma(\omega)]$ is governed by thermal fluctuations, where $E_g$ is the energy gap. In the probe limit gap frequency $\omega_g= 2E_g$ \cite{hartnoll2008holographic}. Now using eq.[\ref{bandgap}] and eq.[\ref{O-Tc1}], we get

\begin{equation}
\frac{\omega_g}{T_c}=\frac{2E_g}{T_c}=7.0894~.
\label{omega_g}
\end{equation}

\noindent Next we evaluate the following ratio using eq.[\ref{O-Tc1}], eq.[\ref{imomega1}], eq.[\ref{rhoTc}] 

\begin{equation}
\lim_{\omega \rightarrow 0}\frac{\omega }{\sqrt{\rho }}Im[ \sigma(\omega)]=0.5583~.
\end{equation}

\noindent This agrees exactly with the numerical result given in \cite{gubser2008gravity}.  

\noindent In this article we also compute the expression of conductivity in self consistent approach and compare the results with our previous estimates. To do that we will be replacing $V$ with its average $\langle V\rangle$ in a self consistent manner. From eq.[\ref{shm1}], we write 

\begin{equation}
A(r_{\ast})= e^{-i\sqrt{\omega^2-\langle V\rangle}~r_{\ast}}.
\label{Ar*}
\end{equation}

\noindent which is consistent with the ingoing boundary condition at the horizon as mentioned earlier.

\noindent From eq.[\ref{sigma}], the expression of conductivity in this case is given by 
\begin{equation}
\sigma(\omega) = \sqrt{1-\frac{\langle V \rangle}{\omega^2}}
\label{sigma2}
\end{equation}

\noindent where,
\begin{equation}
\langle V \rangle = \frac{\int_{-\infty}^{0} V \left|A(r_{\ast})\right|^2 dr_{\ast} }{\int_{-\infty}^{0}\left|A(r_{\ast})\right|^2 dr_{\ast}}~.
\label{aveV}
\end{equation}
\noindent We evaluate the integral considering $\omega$ has an imaginary part, which we will set zero at the end of the calculation.

\noindent Let us now consider a change in variable for better understanding. From eq.[\ref{tor}], we may write
 
 \begin{equation}
 r_{\ast}=-\frac{1}{r_+}\big[z+\frac{z^4}{4}+\frac{z^7}{7}+\frac{z^{10}}{10}+....\big]=-\frac{\tilde{z}}{r_+}~. 
 \label{ztilda}
 \end{equation}
 
 \noindent Now eq.[\ref{aveV}] can be rewritten as following
 
 \begin{equation}
 \langle V \rangle=\frac{\int_{0}^{\infty}V(\tilde{z})e^{-2\sqrt{\langle V \rangle}\sqrt{1-\frac{\omega^2}{\langle V \rangle}}\frac{\tilde{z}}{r_+}}d\tilde{z}}{\int_{0}^{\infty}e^{-2\sqrt{\langle V \rangle}\sqrt{1-\frac{\omega^2}{\langle V \rangle}}\frac{\tilde{z}}{r_+}}d\tilde{z}},~~ V(z)=\frac{\langle\mathcal{O}_{1}\rangle\alpha}{\sqrt{2}}(1-z^3)\tanh^2\left(\frac{b z}{\alpha}\right) ~. 
 \label{aveV1}
\end{equation}
 
 \noindent Notice that at low temperature as $r_+\rightarrow0$, the main contribution to the integral comes when $\tilde{z}\rightarrow0$ and this condition implies $z=\tilde{z}$ or $r_{\ast}(z)=-\frac{z}{r_+}$, which is essentially the region near boundary.
 
 \noindent Hence we put
 \begin{equation}
  V(\tilde{z})=\frac{\langle\mathcal{O}_{1}\rangle\alpha}{\sqrt{2}}(1-\tilde{z}^3)\tanh^2\left(\frac{b\tilde{z}}{\alpha}\right) ~.    
 \end{equation}

\noindent As $\tilde{z}\rightarrow0$, we may write 

\begin{equation}
 \langle V \rangle=\frac{\langle\mathcal{O}_{1}\rangle\alpha}{\sqrt{2}}\frac{\int_{0}^{\infty}\tanh^2(\frac{b\tilde{z}}{\alpha})e^{-2\sqrt{\langle V \rangle}\sqrt{1-\frac{\omega^2}{\langle V \rangle}}\frac{\tilde{z}}{r_+}}d\tilde{z}}{\int_{0}^{\infty}e^{-2\sqrt{\langle V \rangle}\sqrt{1-\frac{\omega^2}{\langle V \rangle}}\frac{\tilde{z}}{r_+}}d\tilde{z}}~.
 \label{avev2}  
\end{equation}
After integrating we deduce 
\begin{eqnarray}
\sqrt{2}\hat{V}=1+2^{\frac{5}{4}}\alpha\sqrt{\hat{V}-\hat{\omega}^2}+2\sqrt{2}\alpha^2(\hat{V}-\hat{\omega}^2)\left[ \mathlarger{\mathlarger{\psi}}\left(\frac{\alpha}{2^\frac{3}{4}}\sqrt{\hat{V}-\hat{\omega}^2}\right)- \mathlarger{\mathlarger{\psi}}\left(\frac{1}{2}+\frac{\alpha}{2^\frac{3}{4}}\sqrt{\hat{V}-\hat{\omega}^2}\right)\right]
\label{aveV3}
\end{eqnarray}

\noindent where

\begin{equation}
  \hat{V}=\frac{\langle V \rangle}{\langle\mathcal{O}_{1}\rangle\alpha},~~~\hat{\omega}^2=\frac{\omega^2}{\langle\mathcal{O}_{1}\rangle\alpha}~. 
  \label{}
  \end{equation}
  
 \noindent For low frequency that is $\omega\rightarrow 0$, from eq.[\ref{aveV3}] we obtain $\hat{V}=0.2924$. Hence for low temperature and low frequency, conductivity given by eq.[\ref{sigma2}] may be written as
 
 \begin{equation}
     \sigma(\omega)=0.489 i \frac{\sqrt{\langle\mathcal{O}_{1}\rangle}}{\omega} ~.
     \label{imomega}
 \end{equation}
 
 \noindent By comparing the above eq.[\ref{imomega}] with the imaginary part of conductivity obtained in eq.[\ref{imomega1}], we see that they are in good agreement.
 
 \noindent At high frequencies that is $\omega\rightarrow\infty$ from eq.(\ref{aveV3}), we obtain $\hat{V}=-\frac{1}{4\alpha^2\hat{\omega}^2}$. Hence at low temperature and high frequency, conductivity given by eq.(\ref{sigma2}) may be written as

\begin{equation}
  \sigma(\omega) = \sqrt{1+\frac{\langle\mathcal{O}_1\rangle^2}{4\omega^4}}~.
  \label{}
\end{equation}
 
\noindent When $\hat{V}$ is comparable with $\hat{\omega}^2$ that is $\hat{V}=\hat{\omega}^2$, from eq.[\ref{aveV3}] we obtain $\langle V \rangle =\frac{\alpha\langle\mathcal{O}_1\rangle}{\sqrt{2}}$ and
\begin{equation}
    \sigma(\omega) = \sqrt{1- \frac{\alpha\langle\mathcal{O}_1\rangle}{\sqrt{2}\omega^2}} ~.
    \label{omega_inter}
\end{equation}
 
\noindent Interestingly there is another way to solve eq.[\ref{shm1}] by treating $\delta V=V-\frac{\alpha\langle\mathcal{O}_1\rangle}{\sqrt{2}}$ as the perturbation and $A(r_\ast)=e^{-i\sqrt{\omega^2-\frac{\alpha\langle\mathcal{O}_1\rangle}{\sqrt{2}}}~r_\ast}$ as zero-th order solution \cite{siopsis2010analytic}. To see this we rewrite eq.[\ref{shm1}] as following
\begin{equation}
 A''(r_{\ast})+\bigg(\omega^2-\frac{\alpha\langle\mathcal{O}_1\rangle}{\sqrt{2}}\bigg) A(r_{\ast})= \bigg(V-\frac{\alpha\langle\mathcal{O}_1\rangle}{\sqrt{2}}\bigg)A(r_\ast)~.   
\end{equation}

\noindent By solving the above equation we obtain 
\begin{eqnarray}
    A(r_\ast)= e^{-i\sqrt{\omega^2-\frac{\alpha\langle\mathcal{O}_1\rangle}{\sqrt{2}}}~r_\ast}
    \Bigg[1+ \frac{\alpha^2}{2\beta}-\frac{\alpha^2 \pi}{\sin\pi\beta}~e^{2i\sqrt{\omega^2-\frac{\alpha\langle\mathcal{O}_1\rangle}{\sqrt{2}}}~r_\ast}+
    \frac{\alpha^2}{2\beta}~{}_2F_{1}\left(1,\beta;1+\beta;-e^{-2\sqrt{\frac{\langle\mathcal{O}_1\rangle}{\sqrt{2}\alpha}}~r_\ast}\right)
    \nonumber\\- \frac{\alpha^2}{2(1+\beta)}e^{-2\sqrt{\frac{\langle\mathcal{O}_1\rangle}{\sqrt{2}\alpha}}~r_\ast}{}_2F_{1}\left(1,1+\beta;2+\beta;-e^{-2\sqrt{\frac{\langle\mathcal{O}_1\rangle}{\sqrt{2}\alpha}}~r_\ast}\right)\Bigg]
    \label{Arstar}
\end{eqnarray}
 
\noindent where  $\beta= i \alpha \sqrt{\frac{\sqrt{2}\omega^2}{\alpha \langle\mathcal{O}_1\rangle }-1}$. Note note that to determine the integration constants we use the ingoing boundary condition at the horizon and we also use the relation ${}_2F_{1}\Big(1,\beta;1+\beta;x\Big)=\frac{e^{i \pi \beta }}{x^\beta}\frac{\pi \beta}{\sin \pi \beta}$ when $\mid x\mid \rightarrow\infty$.

\noindent The expression of conductivity given by  eq.(\ref{sigma}) can be rewritten now as 
\begin{equation}
\label{sigmarstar}
    \sigma(\omega)=\frac{i}{\omega}\left[\frac{\frac{dA(r_{\ast})}{dr_{\ast}}}{A(r_{\ast})}\right]_{r_{\ast}=0}~.
\end{equation}

\noindent Using eq.(\ref{Arstar}) and eq.(\ref{O-Tc1}) in eq.[\ref{sigmarstar}], we deduce 

\begin{equation}
\label{sigmaomega/Tc}
    \sigma(\omega)=\frac{i ~\xi \beta}{2^{\frac{1}{4}}\sqrt{\alpha}\left(\frac{\omega}{T_{c}}\right)}\left[1-\frac{2\left(1+\frac{\alpha^2}{2\beta}\right)}{1-\frac{\pi\alpha^2}{\sin{\pi\beta}}+\frac{\alpha^2}{2}\left\{ \mathlarger{\mathlarger{\psi}}\left(\frac{1+\beta}{2}\right)- \mathlarger{\mathlarger{\psi}}\left(\frac{\beta}{2}\right)\right\}}\right],~~\beta=i\alpha\sqrt{\frac{\sqrt{2}}{\alpha\xi^2}\left(\frac{\omega}{T_{c}}\right)^2-1}~~.
\end{equation}

\begin{figure}[h!]
	\centering
	\includegraphics[width=10cm]{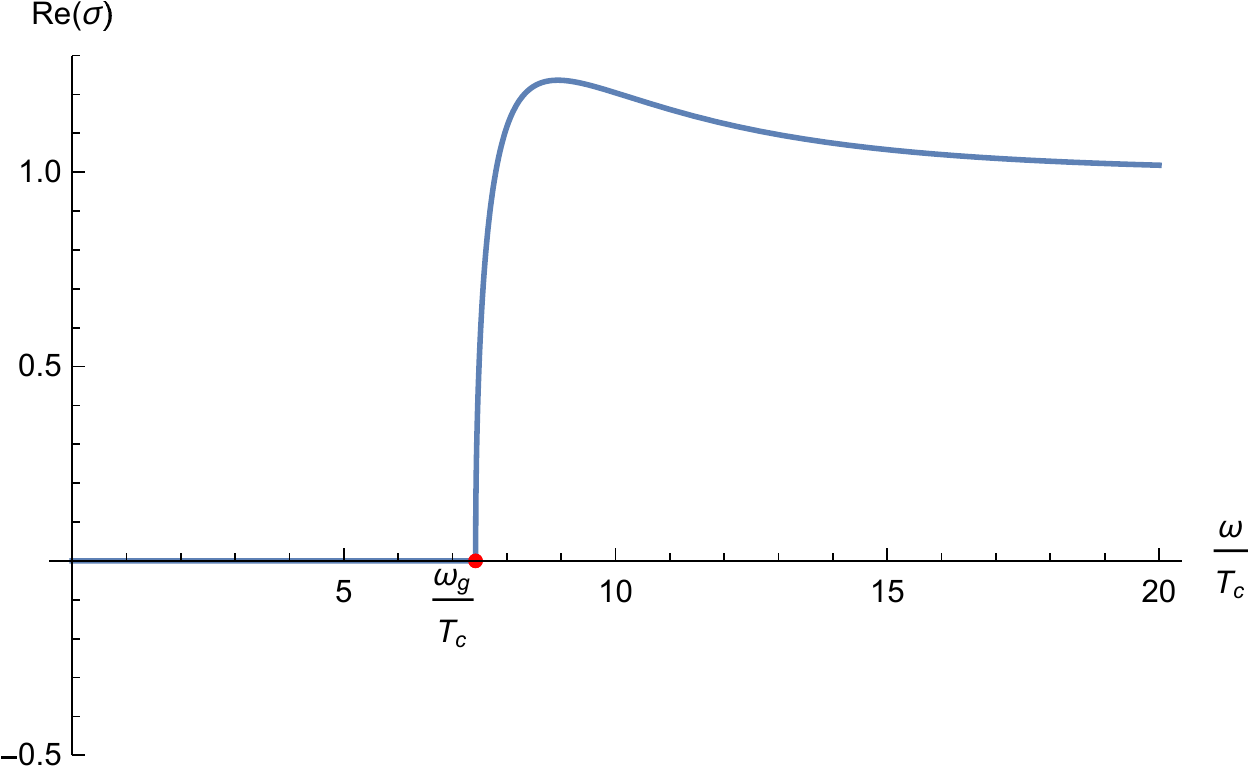}
	
	\caption{$Re(\sigma)$ Vs. $\frac{\omega}{T_c}$ at low temperature}
	\label{re_sigma}
\end{figure}

\begin{figure}[h!]
	\centering
	\includegraphics[width=10cm]{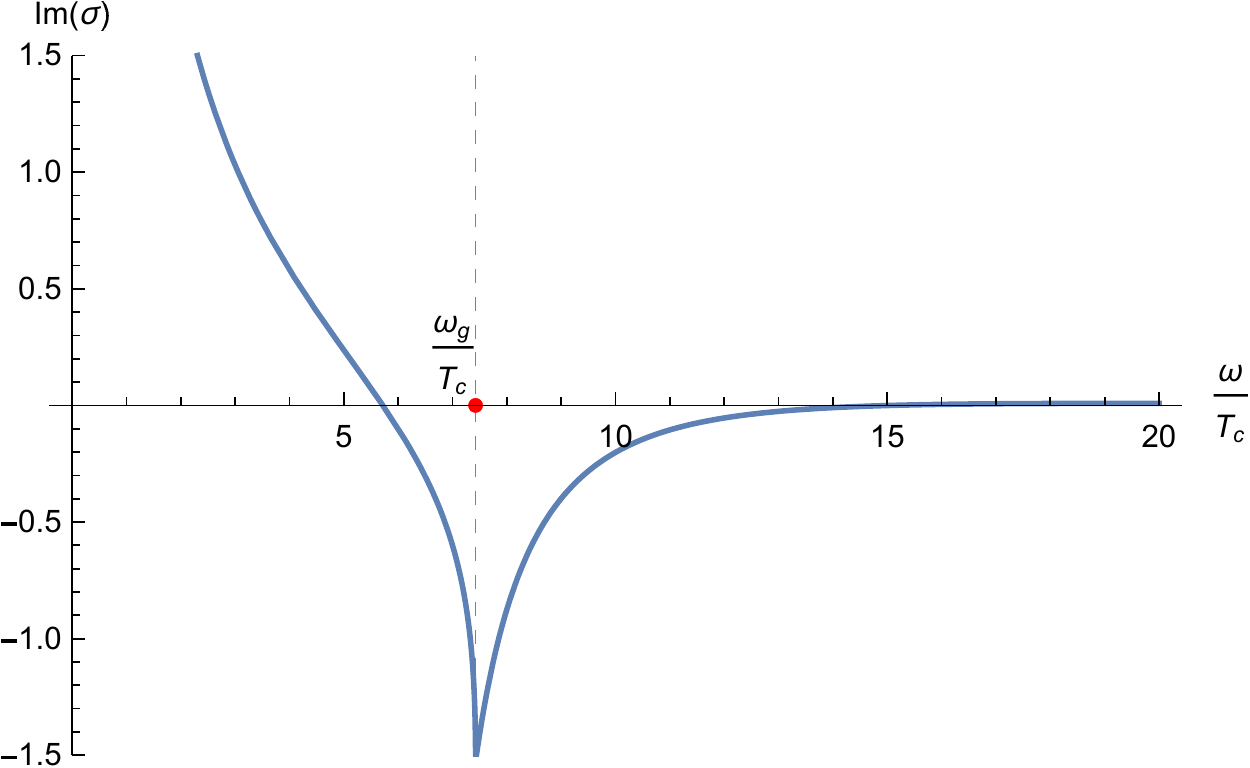}
	
	\caption{$Im(\sigma)$ Vs. $\frac{\omega}{T_c}$ at low temperature}
	\label{im_sigma}
\end{figure}

\noindent We have plotted Figure [\ref{re_sigma}] and Figure[\ref{im_sigma}] using eq.(\ref{sigmaomega/Tc}) which depicts the dependency of $Re[\sigma(\omega)]$  and $Im[\sigma(\omega)]$ on $\frac{\omega}{T_c}$ .

\noindent From Figure[\ref{re_sigma}], we find that at $T\rightarrow 0$, $Re[\sigma(\omega)]$ vanishes for $\omega < \omega_g$ and a gap appears as expected. From the plot we obtain $\frac{\omega_g}{T_c}= 7.4242$ \cite{chen2012effects} which is consistent with our previous estimate given by eq.[\ref{omega_g}].

 \begin{table} [H]
 \footnotesize
\begin{tabular}[t]{|p{3.2cm}|c|}
 \hline
   Approach     &$\frac{\omega_g}{T_c}$  \\
   \hline
   Direct           & 7.0894  \\ 
   \hline
   Perturbative     & 7.424 \\ 
   \hline
 
\end{tabular}
\hfill
\begin{tabular}[t]{|p{3.2cm}|c|}
\hline
   Approach          & $Im[\sigma(\omega)]$  \\
   \hline
    Direct          &0.4616$\frac{\sqrt{\langle\mathcal{O}_{1}\rangle}}{\omega}$ \\ 
   \hline
   Self consistent  & 0.489  $\frac{\sqrt{\langle\mathcal{O}_{1}\rangle}}{\omega}$ \\ 
   \hline
  
\end{tabular}
\hfill
\begin{tabular}[t]{|p{3.2cm}|c|}
\hline
   Approach          & ${}_{\omega\rightarrow0}\frac{\omega }{\sqrt{\rho }}Im[ \sigma(\omega)]$  \\
   \hline
   Analytical  & 0.5583\\ 
   \hline
   Numerical  & 0.55\\ 
   \hline
  
\end{tabular}
\caption{Comparisons for $\frac{\omega_g}{T_c}$, $Im[\sigma(\omega)]$ and ${}_ {\omega\rightarrow0} \frac{\omega }{\sqrt{\rho }}Im[ \sigma(\omega)]$ obtained from different approaches }
\label{t2}
\end{table}

\noindent It is worth mentioning that for $Bi_{2}Sr_{2}CaCu_{2}O_{8+\delta}$ sample, which is a high $T_c$ superconductor, this ratio is found to be  $7.9\pm 0.5$ \cite{gomes2007visualizing}. On the other hand for weakly coupled low $T_c$ superconductors described by BCS theory \cite{bardeen1957theory}, the value of this ratio is $3.5$. Another interesting observation is that at $\omega=0$, there is also a delta function in the $Re[\sigma(\omega)]$ for all $T<T_c$. Although we cannot detect it by analytical or numerical computation as it gives us only the continuous part of $\sigma(\omega)$, this delta function can be revealed by looking at the pole in $Im[\sigma(\omega)]$ at $\omega=0$. A general argument for such a conclusion comes from the Kramers–Kronig relations. Recall that these relations relate the real and imaginary parts of any causal quantity when expressed in frequency space. For conductivity this relation gives us

\begin{equation}
    Im[\sigma(\omega)]= -\frac{1}{\pi} \mathbf{P} \int_{-\infty}^{\infty}\frac{Re[\sigma(\omega')]d\omega'}{\omega'-\omega}~.
\label{kk}
\end{equation}

\noindent From the above relation [\ref{kk}], we conclude that the real part of the conductivity contains a delta function, if and only if the imaginary part has a pole. It is clear from the Figure[\ref{im_sigma}] and from eq.(\ref{imomega1}) there is indeed
a pole in $Im[\sigma(\omega)]$ at $\omega=0$ at low temperature. In the probe limit at $T\rightarrow0$, this delta function in the real part of conductivity implies infinite DC conductivity of superconducting phase. Turning our attention to the gap that appears for the frequencies $\omega<\omega_g$, we conclude there exists a gap in the charge spectrum corresponding to the frequencies $\omega<\omega_g$ and the conduction is non-dissipative. The finite conductivity for $\omega>\omega_g$ indicates dissipation. As $\omega\rightarrow\infty$,  conductivity of the superconducting phase appears to be like the conductivity of the normal phase which in turn implies that the degrees of freedom that contribute to the conductivity at high frequency corresponds to the normal phase.   

\noindent Note that the behaviour of $\sigma(\omega)$ depicted by Fig[\ref{re_sigma}], [\ref{im_sigma}] for p-wave holographic superconductor is qualitatively similar to the $\psi_1=0$ case of s-wave holographic superconductor. The relations  [\ref{imomega1}] and [\ref{bandgap}] in our analysis corresponds to $Im[\sigma(\omega)]=0.55\frac{\sqrt{\langle\mathcal{O}_{2}\rangle}}{\omega}$ and $E_g=0.43\sqrt{\langle\mathcal{O}_{2}\rangle}$ for s-wave holographic superconductor \cite{siopsis2010analytic}.


\section {Conductivity for the case $\psi_1=0$}
Now let us look into the case where we set $\psi_1=0$, and $\psi(z)$ is given by eq.[\ref{psi0}]. Here we simply choose $F(z)=1$ which is consistent with the condition given by eq.[\ref{F(0)}]. In order to deduce conductivity let us substitute eq.(\ref{psi0}) in eq.(\ref{Gz1}) and consider the low temperature rescaleing  $z=\frac{s}{b}$ and by letting $b\rightarrow\infty$, we obtain the following equation
\begin{equation}
\label{G0}
    G^{\prime\prime}(s)+\frac{2 i \omega}{3br_{+}}G^{\prime}(s)-\left[\frac{\langle\mathcal{O}\rangle^2}{2r_{+}^2b^2}-\frac{i \omega}{3r_{+}b^2}-\frac{8\omega^2}{9r_{+}^2b^2}\right]G(s)=0~.
\end{equation}
\noindent Next we choose $b=\frac{\langle\mathcal{O}\rangle}{\sqrt{2}r_{+}}$. Eq.(\ref{G0}) gives the following approximate solution that is valid for low temperature and low frequency $(\omega<<\langle\mathcal{O}\rangle)$ region as
\begin{eqnarray}
    &G(s)&=c_{+}e^{s}+c_{-}e^{-s}\nonumber\\
    \Rightarrow &A(z)&\approx e^{\frac{i\omega z}{3 r_{+}}}\left[c_{+}e^{\frac{\langle\mathcal{O}\rangle}{\sqrt{2}r_{+}}z}+c_{-}e^{-\frac{\langle\mathcal{O}\rangle}{\sqrt{2}r_{+}}z}\right]~.
\end{eqnarray}
 
\noindent Using the definition of conductivity given by eq.(\ref{sigma}), we obtain

\begin{equation}
\label{sigmaforpsi01}
   \sigma(\omega)\approx\frac{i\langle\mathcal{O}\rangle}{\sqrt{2}\omega} \frac{1-\frac{c_{+}}{c{-}}}{1+\frac{c_{+}}{c{-}}}~.
\end{equation}

\noindent  The ratio $\frac{c_{+}}{c_{-}}$ can be found from boundary condition given by eq.(\ref{G(1)-G'(1)}), where we substitute  $\psi(1)\approx\frac{\langle\mathcal{O}\rangle}{\sqrt{2}}$. This gives

\begin{equation}
\label{cpcmpsi0}
    \frac{c_{+}}{c_{-}}=-e^{-2b}\left[\frac{b-3}{b+3}+\frac{2(2b^2-3)}{b(b+3)^2}\frac{i\omega}{r_{+}}+\mathcal{O}(\omega^2)\right]~.
\end{equation}

\noindent By plugging the above ratio in eq.[\ref{sigmaforpsi01}] leads to 

\begin{equation}
\label{sigmaforpsi02}
    \sigma(\omega)=\frac{i\langle\mathcal{O}\rangle}{\sqrt{2}\omega}\left[1+2e^{-2b}\left\{\frac{b-3}{b+3}+\frac{2(2b^2-3)}{b(b+3)^2}\frac{i\omega}{r_{+}}+\mathcal{O}(\omega^2)\right\}\right]~.
\end{equation}

\noindent For the case of low temperature ($T\rightarrow0$) and low frequency ($\omega\rightarrow 0$), this yields the following equations 

\begin{equation}
Im [\sigma(\omega)]\approx\frac{\langle\mathcal{O}\rangle}{\sqrt{2}\omega},~~Re[\sigma(\omega=0)]\sim e^{-2b}\big[1+\mathcal{O}(1/b)\big]=e^\frac{{-E_{g}}}{T}\Rightarrow
E_g=\frac{3}{2\sqrt{2}\pi}\langle\mathcal{O}\rangle\approx0.3376\langle\mathcal{O}\rangle~.
\label{omegapsi01}
\end{equation}

\noindent Next we deduce expression of conductivity using self consistent approach as earlier which is valid for the entire frequency range at low temperature. Here we  substitute $V(z)=\frac{\langle\mathcal{O}\rangle^2}{2}(1-z^3)$ in eq.(\ref{aveV1}) and obtain

\begin{eqnarray}
\label{VO1}
\langle V\rangle=\frac{\langle\mathcal{O}\rangle^2}{2}~.
\end{eqnarray}

 \begin{figure}[h!]
	\centering
	\includegraphics[width=10cm]{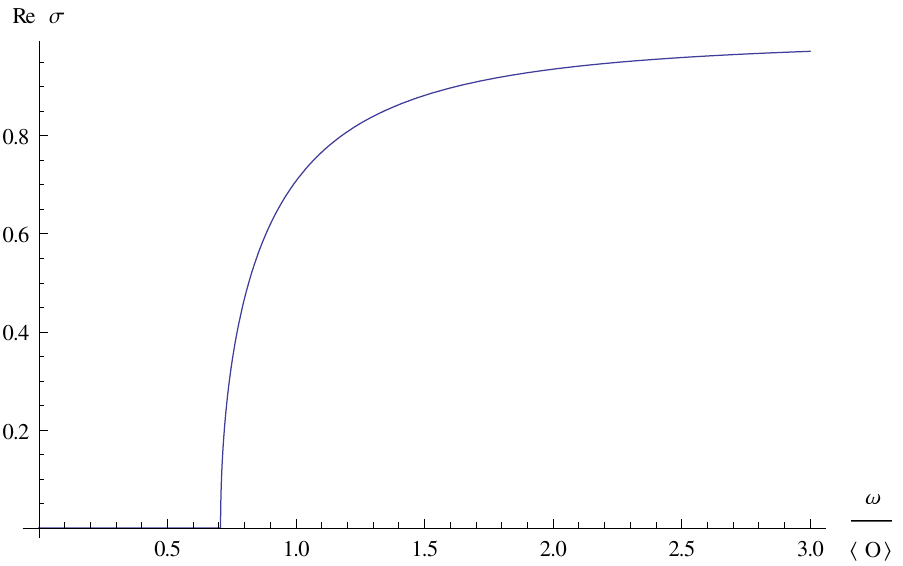}
	
	\caption{$Re(\sigma)$ Vs. $\frac{\omega}{\langle\mathcal{O}\rangle}$ at low temperature}
	\label{im_sigma1}
\end{figure}

\noindent By substituting eq.(\ref{VO1}) in eq.(\ref{sigma2}), we get
\begin{eqnarray}
\sigma(\omega)=\sqrt{1-\frac{\langle\mathcal{O}\rangle^2}{2\omega^2}}~~.
    \label{omegapsi02}
\end{eqnarray}

\noindent For low frequency region, the above expression leads to 

\begin{equation}
  Im [\sigma(\omega)]=\frac{\langle\mathcal{O}\rangle}{\sqrt{2}\omega}~. 
   \label{omegapsi03}
\end{equation}

\noindent This matches perfectly with our previous estimate given by eq.[\ref{omegapsi01}]. In Figure[\ref{im_sigma1}], we have also plotted $Re[\sigma(\omega)]$ vs. $\frac{\omega}{\langle\mathcal{O}\rangle}$ using eq.[\ref{omegapsi02}].
Note that the above analysis ( $\psi_1=0$ case of p-wave holographic superconductor) is qualitatively similar to that of $\psi_2=0$ case of s-wave holographic superconductor and eq.[\ref{omegapsi01}] for p-wave corresponds to $Im[\sigma(\omega)]=\frac{\langle\mathcal{O}_{1}\rangle}{\omega}$ and $E_g=0.48\langle\mathcal{O}_{1}\rangle$ for s-wave holographic superconductor \cite{siopsis2010analytic}.

\section{Conclusions}
This paper mostly focuses on the analytical computation of the conductivity of p-wave holographic superconductors described by Einstein-Yang-Mills theory in the probe limit. In section 3 the system was analysed around critical temperature ($T_c$) above which the condensate vanishes. We obtained the behavior of the field variables $\phi$ and $\psi$. Established the relationship between critical temperature and charge density ($\rho$). In section 4 we have  discussed the approximate behavior of field variables $\psi$ and $\phi$ by solving the coupled field equations analytically by two different approaches at low temperature limit. We have also provided the relationship between condensation operator $\langle\mathcal{O}_1\rangle$ and $T_c$ at low temperature.

 In section 5 we have discussed the conductivity for the case where $\psi_0$ is set to zero. First we derived the expression of conductivity at low frequency and low temperature and established the fact that at low frequency limit ($\omega\rightarrow0$) the real part of $\sigma(\omega)$ is governed by thermal fluctuations as  $\lim_{\omega \rightarrow 0} Re[\sigma(\omega)]\sim e^{\frac{-E_g}{T}}$ and computed the value of the ratio $\frac{\omega_g}{T_c}$. We also obtained the expression of conductivity for the entire frequency range using self-consistent approach. Then using perturbation techniques the field equation [\ref{shm1}] for the gauge field $A$ was solved. In Figures [\ref{re_sigma}] and [\ref{im_sigma}], we showed the dependency of the real and imaginary part of conductivity on frequency at low temperature limit and also obtained the ratio $\frac{\omega_g}{T_c}$ from the plots which is consistent with the previously obtained result. 

 Next in section 6, we have computed the expression for conductivity for the case where $\psi_1$ is set to zero. Another interesting observation we made is that at low temperature the gap energy $E_g$ is proportional to $\sqrt{\langle\mathcal{O}_1\rangle}$ if we consider the conformal dimension one, that is $\psi_0=0$ case, and proportional to $\langle\mathcal{O}\rangle$ if we consider the conformal dimension zero that is, $\psi_1=0$ case \cite{hartnoll2008holographic}.

\section*{Appendix}
Here we will derive the expression for holographic conductivity eq.(\ref{sigma}). Matter Lagrangian is given by
\begin{eqnarray}
S_{m}&=&-\frac{1}{4}\int d^4x\sqrt{-g}F^a_{\mu\nu}F^{a\mu\nu}\nonumber\\
&=&-\frac{1}{4}\int d^4x\sqrt{-g} \left[\nabla_{\mu}A^a_{\nu}-\nabla_{\nu}A^a_{\mu}+q f^{abc}A^{b}_{\mu}A^{c}_{\nu}\right]F^{a\mu\nu}\nonumber\\
&=&-\frac{1}{2}\int d^4x\sqrt{-g}\nabla_{\mu}\left(F^{a\mu\nu}A^a_{\nu}\right)+\frac{1}{2}\int d^4x\sqrt{-g} \nabla_{\mu}\left(F^{a\mu\nu}\right)A^a_{\nu}-\frac{q}{4}\int d^4x\sqrt{-g}f^{abc}A^{b}_{\mu}A^{c}_{\nu}F^{a\mu\nu}~.\nonumber\\
\end{eqnarray}

\noindent Now using eq.(\ref{eom1}), the on shell action is given by

\begin{eqnarray}
S_{o.s}&=&-\frac{1}{2}\int d^4x\sqrt{-g}\nabla_{\mu}\left(F^{a\mu\nu}A^a_{\nu}\right)+\frac{q}{4}\int d^4x\sqrt{-g}f^{abc}A^{b}_{\mu}A^{c}_{\nu}F^{a\mu\nu}\nonumber\\
&=&-\frac{1}{2}\int_{\partial M}d^3x\sqrt{-h}n_{\mu}F^{a\mu\nu}A^a_{\nu}+\frac{q}{4}\int d^4x\sqrt{-g}f^{abc}A^{b}_{\mu}A^{c}_{\nu}F^{a\mu\nu}~~.
\label{Sos}
\end{eqnarray}

\noindent Using the ansatz $A^3_t=\phi(r)$, $A^1_x=\psi(r)$ and $A^3_y=\delta A_y$ in eq.(\ref{Sos}), we will get the on shell action to be

\begin{eqnarray}
S_{o.s}&=&-\frac{1}{2}\int d^3x\left[f(r)A^3_y\partial_rA^3_y+f(r)\psi(r)\psi^{\prime}(r)-r^2\phi(r)\phi^{\prime}(r)\right]_{r\rightarrow\infty}-\frac{q^2}{2}\int d^4x\frac{\psi^2(r)\phi^2(r)}{r^2f(r)}\nonumber\\
&+&\frac{q^2}{2}\int d^4x\frac{\psi^2(r)}{r^4}(\delta A_{y})^2~~.
\label{Sos1}
\end{eqnarray}

\noindent We can neglect the last term of eq.(\ref{Sos1}) since it contains the perturbation square term. Asymptotic behaviour of the perturbation field is (from eq.(\ref{bc3}))
\begin{equation}
A^{3}_{y}=A^{(0)}+\frac{A^{(1)}}{r}+...~.
\end{equation}
According to the AdS/CFT correspondence, the
electrical current based on the on-shell bulk action $S_{o.s}$ reads

\begin{equation}
J_y=\frac{\delta S_{o.s}}{\delta A^{(0)}}=A^{(1)}.
\end{equation}

\noindent In the last line to compute variation of $S_{o.s}$ with respect to $A^{(0)}$, we use the fact that $A^{(1)}$ is proportional to $A^{(0)}$\cite{hartnoll2008holographic}. Electric field at boundary is given by $E_y=-[\partial_t(\delta A_y)]_{r\rightarrow\infty}$. So conductivity is given by
\begin{equation}
\sigma_{yy}=\frac{J_{y}}{E_y}=-\frac{i A_1}{\omega A_0}~~.
\end{equation}

\bibliographystyle{ieeetr}
\bibliography{ref.bib}


\end{document}